\documentclass[12pt,preprint]{aastex}
\usepackage{graphicx}
\usepackage{epsfig}
\usepackage{multirow}
\usepackage{footnote}
\usepackage{amsmath}
\usepackage{amssymb}

\shorttitle{Dust around HD\,131488 and HD\,121191}
\shortauthors{C. Melis et al.}

\usepackage{color}
\usepackage{graphicx}

\begin{document}


\title{Copious Amounts of Hot and Cold Dust Orbiting the Main Sequence A-type Stars HD\,131488 and HD\,121191}


\author{Carl Melis\altaffilmark{1,8}, B. Zuckerman\altaffilmark{2}, Joseph H. Rhee\altaffilmark{3}, Inseok Song\altaffilmark{4}, Simon J. Murphy\altaffilmark{5,6,7}, Michael S. Bessell\altaffilmark{5}}
\email{cmelis@ucsd.edu}


\altaffiltext{1}{Center for Astrophysics and Space Sciences, University of California, San Diego, CA 92093-0424, USA}
\altaffiltext{2}{Department of Physics and Astronomy, University of California,
Los Angeles, CA 90095-1547, USA}
\altaffiltext{3}{Department of Physics and Astronomy, California State Polytechnic University, Pomona, 3801 W. Temple Ave., Pomona, CA 91768, USA}
\altaffiltext{4}{Department of Physics and Astronomy, University of Georgia, Athens, GA 30602-2451, USA}
\altaffiltext{5}{Research School of Astronomy and Astrophysics, Australian National University, Canberra, ACT 2611, Australia}
\altaffiltext{6}{Current address: Astronomisches Rechen-Institut, Zentrum f\"{u}r Astronomie der Universit\"{a}t Heidelberg, Germany 69120}
\altaffiltext{7}{Gliese Fellow}
\altaffiltext{8}{Joint NSF AAPF Fellow and CASS Postdoctoral Fellow}


\begin{abstract}
We report two new dramatically dusty main sequence stars: HD\,131488 (A1\,V) and 
HD\,121191 (A8\,V). HD\,131488 is found to have substantial amounts of dust in its 
terrestrial planet zone (L$_{\rm IR}$/L$_{\rm bol}$$\approx$4$\times$10$^{-3}$), 
cooler dust further out in its planetary system, and an unusual mid-infrared spectral feature. 
HD\,121191 shows terrestrial planet zone dust
(L$_{\rm IR}$/L$_{\rm bol}$$\approx$2.3$\times$10$^{-3}$),
hints of cooler dust, and shares the unusual mid-infrared 
spectral shape identified in HD\,131488. These two stars belong to sub-groups of the 
Scorpius-Centaurus OB association and have ages of $\sim$10\,Myr.
HD\,131488 and HD\,121191 are the dustiest main sequence A-type stars currently known.
Early-type stars that host substantial inner planetary 
system dust are thus far found only within the age range of 5-20\,Myr.
\end{abstract}





\section{Introduction}
\label{secint}

With {\it Infrared Astronomical Satellite} ({\it IRAS}; \citealt{neugebauer84}),
$AKARI$ (\citealt{murakami07}), $Spitzer$ $Space$ $Telescope$ ($Spitzer$; \citealt{werner04}), $Wide$ $Field$ $Infrared$ $Survey$ $Explorer$ ({\it WISE}; \citealt{wright10}), and ground-based
data our team has discovered 
and characterized some of the dustiest main sequence star systems currently 
known (e.g., BD+20~307, \citealt{song05}; EF\,Cha, \citealt{rhee07b}; 
HD\,23514, \citealt{rhee08}; HD\,15407, \citealt{melis10}; 
V488\,Per, \citealt{zuckerman12}; 
TYC\,8241~2652~1, \citealt{melis12}; HD\,166191, \citealt{schneider13}). 
These discoveries are part of our ongoing 
effort to build observational evidence in support of the collisional formation model
for rocky planets like Earth around stars of various masses
(e.g., \citealt{melis10}, Melis {\it et al}.\ 2013 in preparation). 
From these systems we are able to determine that, for terrestrial-like planets, the epoch of final 
mass accumulation through giant impacts occurs at ages between 30 and 100\,Myr for stars of 
roughly Solar mass \citep{melis10} 
and ages of 5-20\,Myr for stars a few times the mass of the Sun (see Section \ref{secdisc}).

This paper reports two new dusty A-type stars discovered in our ongoing
search of mid-infrared databases. These two systems are the dustiest
A-type stars discovered to date. Section \ref{secobs} discusses observational follow-up of these
systems, Section \ref{secres} reports results and analysis on all available data, and discussion
is provided in Section \ref{secdisc}.


\section{Observations}
\label{secobs}

HD\,131488 was identified as a potentially dusty star 
in a cross-correlation of the Tycho-2 \citep{hog00} catalog of
stellar sources and the
$IRAS$ Faint Source Catalog (\citealt{moshir92}; 
more details about this cross-correlation can be found
in \citealt{melis09phd}). HD\,121191 was identified in a cross-correlation of Tycho-2
with the $AKARI$ InfraRed Camera catalog (\citealt{onaka07}; \citealt{ishihara10}).
Ground-based observations for these candidate dusty stars were conducted to verify the 
suggestion of excess emission and to characterize the stellar hosts.

\subsection{Imaging}

\subsubsection{Mid-infrared}
\label{secirim}

Ground-based follow-up imaging in the mid-infrared was pursued at 
Gemini-South with T-ReCS \citep{telesco98}. T-ReCS 
hosts a 320$\times$240 pixel array whose 0.09$^{\prime\prime}$\,pixel$^{-1}$ plate scale
affords a field of view of 28.8$^{\prime\prime}$$\times$21.6$^{\prime\prime}$.
Observations were performed with the default chop-nod parameters:
15$^{\prime\prime}$ chop throws along a position angle of 0$^{\circ}$ and
an ABBA nod pattern.
Data reduction followed standard high thermal background techniques. Chop pairs are 
differenced to remove the rapidly fluctuating background 
signal. Stacked chop pairs of different nods are then 
combined to yield the final reduced image. Observations were
performed with the N-band filter
(broad 10\,$\mu$m for T-ReCS; see Table \ref{tabirim}) 
and aimed to verify {\it IRAS} and/or {\it AKARI} $\sim$10\,$\mu$m-detected 
flux levels and to rule out contamination by nearby sources. Flux
calibration standards were selected from lists of well-characterized, mid-infrared
bright stars (e.g., \citealt{cohen99}). Mid-infrared observations
are summarized in Table \ref{tabirim} and retrieved flux densities shown in
Table \ref{tabphot} and Figure \ref{fig121191sed}. Although the epoch 2009 images
of HD\,131488 were performed in non-photometric conditions, a 
point-spread function calibration star was observed
to enable a search for extended emission toward HD\,131488 in either 
band-pass.

\subsubsection{Adaptive Optics}

L$'$-band (3.776$\pm$0.7\,$\mu$m) adaptive optics imaging of HD\,131488 was performed
on UT 17 July 2009 to
further search for small-separation contaminating sources. Observations
were carried out with the Mauna Kea Observatory Keck Adaptive Optics system
\citep{wizinowich00}. The Keck Adaptive Optics
system was fed into NIRC2, a camera with a 
1024 $\times$ 1024 InSb Aladdin-3 array. Observations were performed
in the ``Narrow'' camera mode with a plate scale of $\approx$0.01$''$ pixel$^{-1}$.
HD\,131488 served as the guide star for
the adaptive optics systems and provided reasonable atmospheric correction
(Strehl ratio $\gtrsim$0.4). 
Data obtained for HD\,131488 did not
make use of a dither sequence; when reducing the data 
background subtraction is done with
a subsequent image of a field close in the plane of the sky to HD\,131488.
Individual images are then flat-fielded, rotated such that North points up in each image, and 
stacked to yield the final reduced image.

\subsection{Spectroscopy}
\label{secspec}

Spectroscopic observations were performed in the optical as well as in
the infrared to characterize host stars and their infrared excess.

\subsubsection{Optical}

Optical echelle and grating spectroscopy was carried out at
Siding Spring Observatory (SSO) with the 2.3-m telescope and at Mauna
Kea Observatory with the Keck~I 10-m telescope.
High-spectral resolution data were obtained with
HIRES \citep{vogt94} at Keck and the SSO 2.3-m
echelle (SSO-E).
Moderate-resolution grating spectroscopy were obtained with the
Double-Beam Spectrograph (DBS; \citealt{rodgers88}) and WiFeS \citep{dopita07}
at SSO.
Observation parameters $-$ including wavelength range covered,
instrumental resolution, and resulting signal-to-noise ratio of the spectrum $-$ can be found in 
Table \ref{taboptspec}.

HIRES data are reduced and extracted with the {\sf MAKEE}
software package. 
Siding Spring (except WiFeS)
data are reduced following standard IRAF tasks and procedures.
WiFeS integral-field unit data were obtained in single-star mode with twice 
the spatial binning (1$''$ spatial pixels). Spectra are obtained by optimally extracting
and combining five image slices (effectively a 5$''$ diameter
aperture around the object) that contain the majority of the stellar flux. Low-resolution
optical spectra for HD\,131488 and HD\,121191 are shown in Figure \ref{fig121191wifes}.

\subsubsection{Infrared}

Infrared spectroscopic
observations sought to further characterize the slope of the continuum excess emission
and to identify solid-state emission features.

\begin{figure}
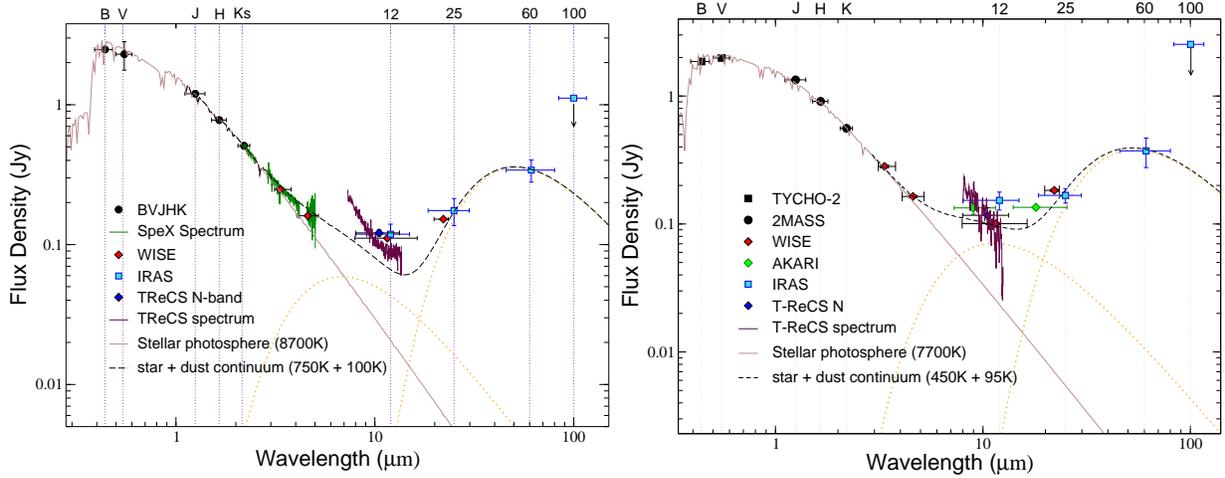

 \centering
 \begin{minipage}[t!]{80mm}
  \includegraphics[width=80mm]{figure1a.eps}
 \end{minipage}
 \begin{minipage}[t!]{80mm}
  \includegraphics[width=80mm]{figure1b.eps}
 \end{minipage}
 \caption{\label{fig131488sed} \label{fig121191sed} Spectral energy distributions for
                HD\,131488 and HD\,121191. The data points blueward of 1\,$\mu$m are $BV$ 
                measurements from Tycho-2. The circle
                data points redward of 1\,$\mu$m are $JHK_{\rm s}$ measurements from 
                2MASS.
                Horizontal lines in the data points indicate filter
                band-passes while the vertical lines in data points indicate the measurement
                uncertainty. Some measurement uncertainties are smaller
                than the point sizes on the plot. The vertical scalings of T-ReCS spectra 
                are calibrated with the T-ReCS N-band imaging
                data point.
                The solid brown curve connecting the $BVJHK_{\rm s}$ data points
                is a synthetic stellar atmospheric spectrum \citep{hau99}. 
                The dashed orange curves are blackbodies at the temperatures indicated
                on the figure panel. The dashed curve is the sum of
                the atmospheric and blackbody models.
                }
\end{figure}

\begin{figure}
 \centering
 \begin{minipage}[t!]{80mm}
  \includegraphics[width=90mm]{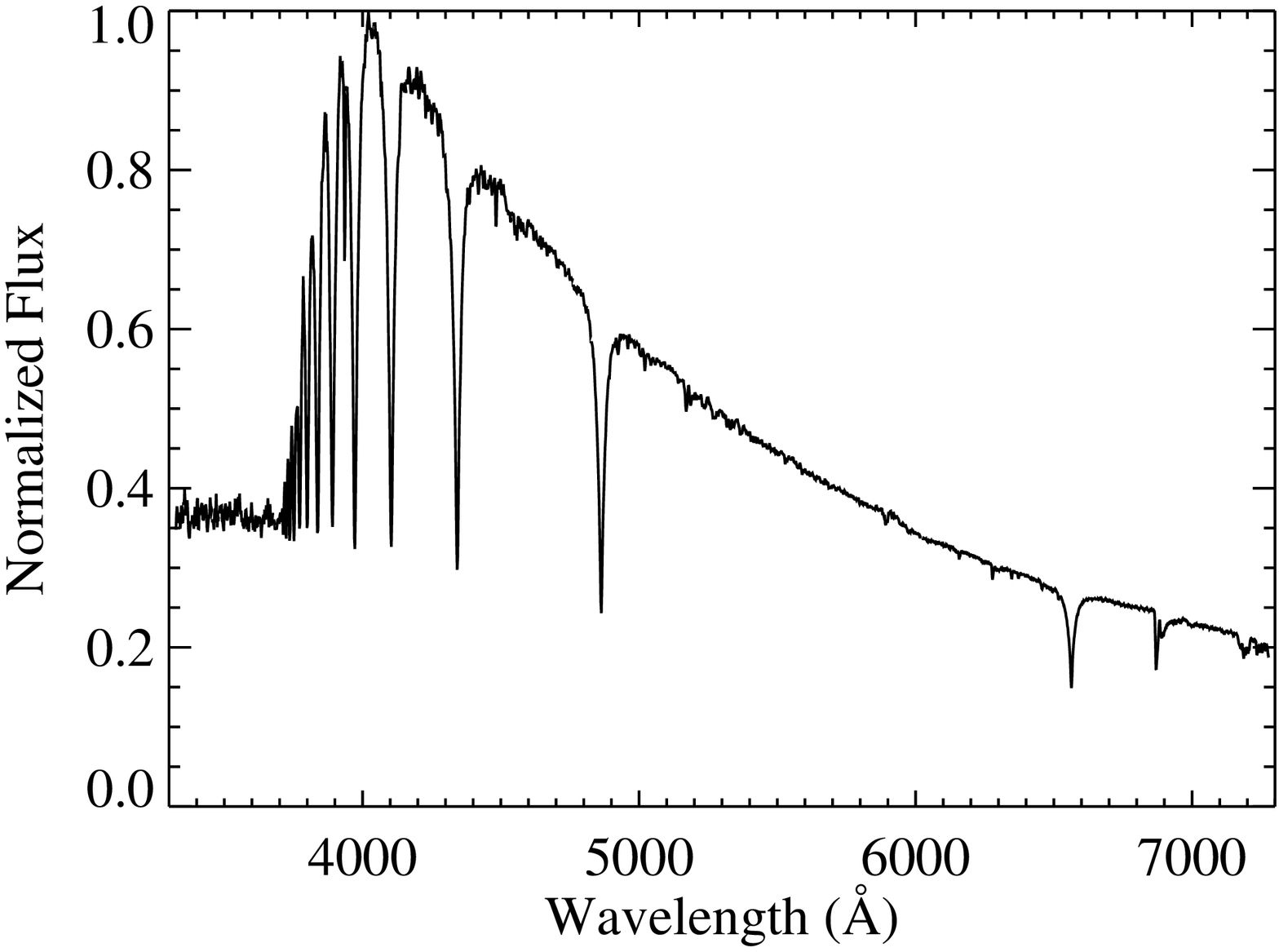}
 \end{minipage}
 \begin{minipage}[t!]{80mm}
  \includegraphics[width=90mm]{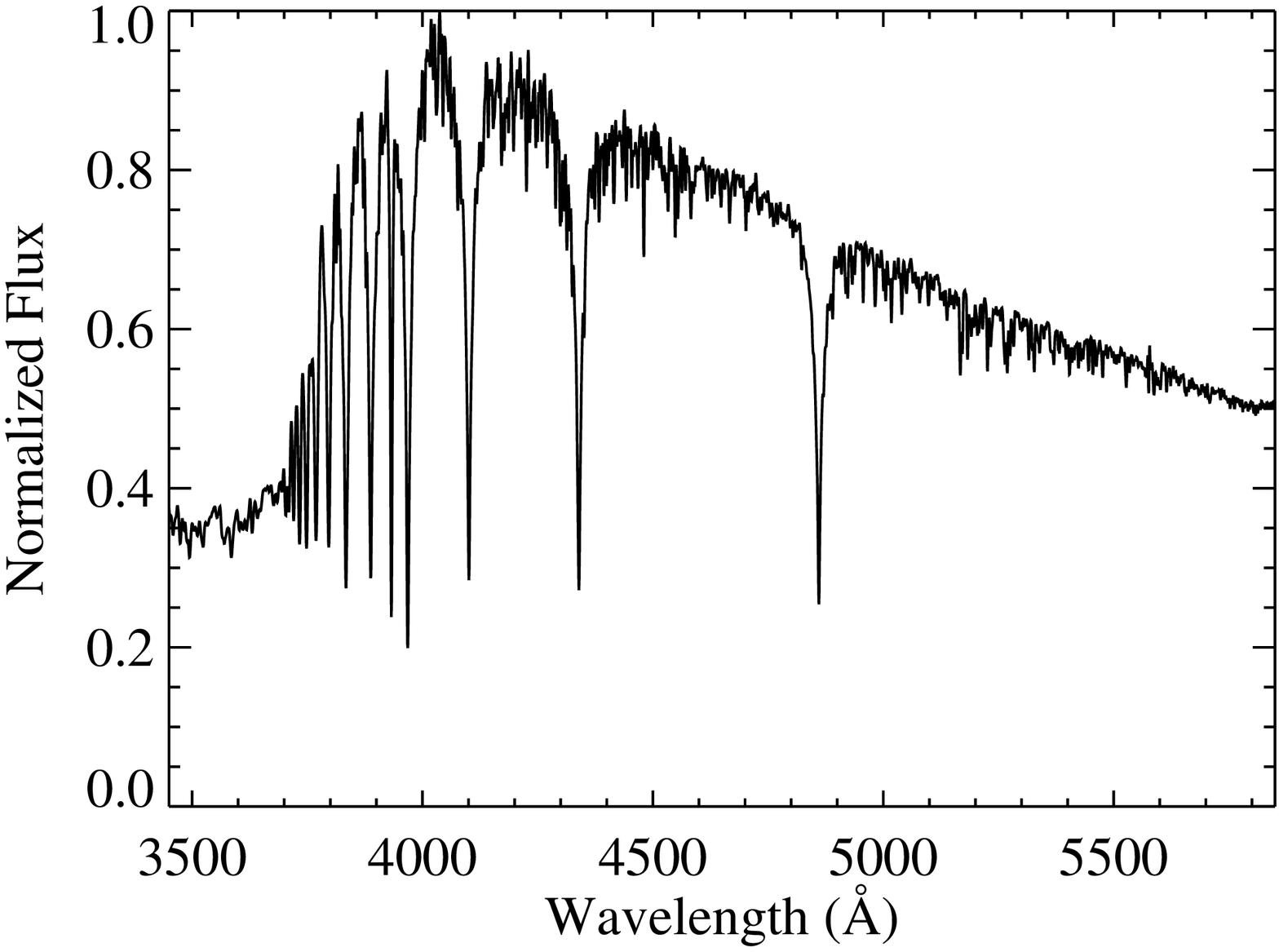}
 \end{minipage}
 \caption{\label{fig131488dbs} \label{fig121191wifes} {\it Left:} Complete DBS spectrum of 
                HD\,131488.
                {\it Right:} WiFeS-blue spectrum of HD\,121191. 
                The wavelength
                scale in these figures is in air.
                There are no noticeable emission features nor any obvious filling of the
                Balmer absorption lines in these spectra nor in the high-spectral resolution
                data reported in Table \ref{taboptspec}. This indicates that these stars are unlikely to
                be actively accreting gas-rich material and, along with arguments provided in
                Section \ref{secdisc}, shows that their orbiting dust is second generation
                material produced in the collisions of mature rocky objects.
                }
\end{figure}

Ground-based mid-infrared spectroscopy was pursued at
Gemini-South with T-ReCS. 
The observing strategy was similar to that used for imaging (Section \ref{secirim}),
except a slit and grating were employed. Slit sizes and
other spectroscopic parameters are listed in Table \ref{tabirspec}.
Data are reduced in the same way as imaging data (Section \ref{secirim}).
Spectral data are then extracted using in-house software routines.
Telluric features are removed by dividing target spectral data by spectral 
data of a standard star. Observational details, including standard stars, can
be found in Table \ref{tabirspec}. Mid-infrared spectra for both stars can be found
in Figures \ref{fig121191sed} and \ref{fig131488speccomp}.


IRTF SpeX \citep{rayner03} observations of HD\,131488 were performed 
during engineering time; observation parameters can be found in Table \ref{tabirspec}.
The SpeX data for HD\,131488 were obtained at an average
airmass of 2.15 (close to transit for Mauna Kea). Spectral
data are reduced, extracted, wavelength calibrated, and telluric calibrated using 
SpeXTool \citep{cushing04,vacca03}. Telluric features in the SpeX data are
removed by dividing the target spectrum by a standard star (HD\,130163)
observed at similar airmass using a similar observation sequence. The final reduced,
flux calibrated spectrum is shown in Figure \ref{fig121191sed}.




\section{Results and Analysis}
\label{secres}

Here observations are synthesized to characterize each host star and its orbiting dust.
In addition to the observations reported above, photometric measurements for each
source from Tycho-2, 2MASS \citep{skrutskie06}, $IRAS$, $AKARI$, and $WISE$
are collected and reported in Table \ref{tabphot}. No archival $Spitzer$ observations
exist for either HD\,131488 nor HD\,121191.

\subsection{HD\,131488}
\label{sec131488}

\subsubsection{Association and Age of HD\,131488}
\label{sec131488age}

HD\,131488 is located within the greater Scorpius-Centaurus OB association
(ScoCen; e.g., \citealt{preibisch08}) and resides in a region of the sky
that is typically ascribed to the Upper-Centaurus-Lupus association
\citep[UCL; see Table \ref{tabpars} herein and Figure 8 of][]{zuckerman04}.
Age estimates for the UCL range from 10-20\,Myr 
(\citealt{degeus89,dezeeuw99,sartori03,song12}). 
It is interesting to note that if HD\,131488 has such a young age,
then it should be on the zero-age main sequence for A-type stars and
lie with other young A-type stars on the HR-diagram
presented in Figure 5 of \citet{zuckerman01}. 
Under this consideration it is possible to make estimates for the
distance to HD\,131488 with the assumption that it is young. These
estimates can then be checked for robustness by seeking agreement
with kinematics of the UCL and HD\,131488 (indeed, as shown below,
such agreement is found). For distances of 190\,pc, HD\,131488 would
appear close to the Pleiades A-type main sequence, while for distances less
than 150\,pc it would appear below the ZAMS for A-type stars. Thus,
to have agreement with the zero-age
main sequence for A-type stars, a distance estimate of
150\,pc (and a distance range of 150-180\,pc) is obtained for HD\,131488.
Such distances agree with those for the UCL \citep{dezeeuw99}.

We obtained several high- and medium-resolution optical echelle spectra of
HD\,131488 to determine its radial velocity (and hence kinematics to compare to
those of the UCL) and whether or not
the radial velocity varied (to search for evidence of binarity). 
No evidence for radial velocity variation
is seen between individual epochs to within the measurement errors (which
are typically $\approx$3\,km s$^{-1}$ for this rapidly rotating A-type star). The
radial velocity quoted in Table \ref{tabpars} is the average of all five
epochs.
Combining this measurement with the Tycho-2 measured proper
motion and the above photometric distance estimate
allows us to calculate the UVW space motions for 
HD\,131488\footnote{UVW are defined with respect to the Sun; we employ a 
right-handed coordinate system and have positive U towards the Galactic center, positive V 
in the direction of Galactic rotation, and positive W toward the North Galactic pole.}.
Comparing HD\,131488's UVW space motions (Table \ref{tabpars})
to those of the UCL (U$=-$5.1$\pm$0.6\,km\,s$^{-1}$, V$=-$19.7$\pm$0.4\,km\,s$^{-1}$, 
W$=-$4.6$\pm$0.3\,km\,s$^{-1}$; \citealt{sartori03,chen11})
secures our identification of HD\,131488 as a member of that association.
We adopt an age of 10\,Myr for HD\,131488, consistent with the considerations
of \citet{song12}. 

\subsubsection{Characterizing HD\,131488's Infrared Excess}
\label{sec131488irex}

T-ReCS N-band photometric imaging recovers 
a point source having flux consistent
with the {\it IRAS} 12\,$\mu$m data point. Further imaging at N- and Qa-band
allowed us to identify, to within 0.5$^{\prime\prime}$, HD\,131488
as the carrier of the mid-infrared excess. No extension in the point-spread function of any
mid-infrared imaging was detected.
$L'$-band
adaptive optics imaging also recovers a single point source
with a core point-spread function full-width at half-maximum 
of $\approx$80\,milliarcseconds.

SpeX thermal-infrared spectroscopy in the 2-5\,$\mu$m range 
suggests that the excess emission towards HD\,131488
begins at wavelengths as short as $\approx$4\,$\mu$m (Figure \ref{fig131488sed}).
T-ReCS N-band spectroscopy reveals a strange profile reminiscent of
the Rayleigh-Jeans slope of a blackbody (Figure \ref{fig131488sed}). 
However, the flux level of the spectrum is significantly above the stellar photosphere. 
Since it is unlikely that there is a contaminating background source, this
strange profile is attributed to material associated with the star. The possibility of
a solid-state feature is discussed in Section \ref{secdisc}.

We estimate the dust temperature and fractional infrared
luminosity of the material surrounding HD\,131488 through model fitting. 
We fit optical and
near-infrared measurements out to J-band with a synthetic stellar
atmosphere spectrum \citep{hau99} to estimate the stellar flux
and add blackbodies (eventually found through $\chi$$^2$ minimization to have
temperatures of 750 and 100\,K) to fit the near- to far-infrared
excess emission. Two blackbodies are used as a single blackbody cannot reasonably
fit both the short- and long-wavelength excess emission. Decomposing the dust into two
components, hot and cold, we estimate that the cool component
contributes $\approx$2$\times$10$^{-3}$
to the total fractional infrared luminosity 
(total L$_{\rm IR}$/L$_{\rm star}$$\approx$6$\times$10$^{-3}$) 
while the hot component contributes
$\approx$4$\times$10$^{-3}$. From the fitted temperatures, and adopting the
spectral energy distribution fit
T$_{\rm star}$=8800\,K and the derived L$_{\rm star}$=15.5\,L$_{\odot}$, we 
determine that blackbody dust grains having temperatures of 750 and 100\,K reside
at  $\approx$0.6\,AU and $\approx$35\,AU separations from HD\,131488, respectively.
We note that this temperature decomposition is ambiguous due to the incomplete
far-infrared coverage of the cool dust excess emission. Far-infrared measurements
extending beyond 60\,$\mu$m will enable an unambiguous decomposition of the 
hot and cold dust regions and perhaps further enable identification of the strange 
absorption/emission feature seen in the T-ReCS spectrum. 

\subsection{HD\,121191}

\subsubsection{Association and Age of HD\,121191}

HD\,121191 falls in the plane of the sky between the UCL and
Lower-Centaurus-Crux associations (LCC; e.g., see Table \ref{tabpars} and
Figure 8 of \citealt{zuckerman04}). Using a methodology similar to that
described for HD\,131488 (Section \ref{sec131488age}), we estimate
a distance to HD\,121191 of roughly 130\,pc and a distance
range of 120-140\,pc. Such a distance is in agreement with accepted distances
to UCL and LCC \citep[e.g.,][]{dezeeuw99}. Two epochs of SSO-E measurements
reveal no evidence for radial velocity variability and the value
reported in Table \ref{tabpars} is the average of both epochs. We compute UVW
space motions from the obtained kinematic data and find
that HD\,121191 has space motions consistent with either of the
UCL or LCC sub-regions \citep[Table \ref{tabpars}; LCC has 
U$=-$7.8$\pm$0.6\,km\,s$^{-1}$, V$=-$20.7$\pm$0.6\,km\,s$^{-1}$, 
W$=-$6.0$\pm$0.3\,km\,s$^{-1}$; e.g.,][]{sartori03,chen11}.
Similar to the UCL sub-region, age estimates for the LCC
range from 10-20 Myr \citep{dezeeuw99,sartori03,pecaut12,song12}.
We adopt an age of 10\,Myr for HD\,121191.

\subsubsection{Dust Orbiting HD\,121191}
\label{sec121191irex}

T-ReCS N-band imaging recovers flux consistent with the {\it AKARI}
measurement and shows that there is no contaminating source outside
of $\sim$0.4$''$ of HD\,121191.
Measurements from the {\it IRAS} Faint Source Reject Catalog
suggest strong mid- and far-infrared excess emission toward HD\,121191 (Table \ref{tabphot}
and Figure \ref{fig121191sed}).
However, the {\it IRAS} catalog entry for HD\,121191 states that there are 
five cirrus-only extractions near its location and examination of the $IRAS$ 
Scan Processing and Integration tool 
(ScanPI\footnote{http://scanpi.ipac.caltech.edu:9000/applications/Scanpi/index.html}) 
shows a strong 60\,$\mu$m detection at the position of HD\,121191 cast over a strongly
sloping and structured background. This is suggestive of significant Galactic cirrus
(interstellar medium dust clouds) near the location of HD\,121191 that could be generating 
a Pleiades-type effect \citep{kalas02}. Although the T-ReCS measurements provide
evidence against such an interpretation for the mid-infrared excess emission
(such emission would be resolved by T-ReCS; see the case of 29\,Per
in \citealt{zuckerman12}), it is not clear that the far-infrared excess emission
also originates from circumstellar dust and not from cirrus. {\it WISE} 22\,$\mu$m
measurements of the star show no evidence of extent and report flux density
consistent with the {\it IRAS} 25\,$\mu$m measurement, hinting that the
60\,$\mu$m flux may indeed originate from circumstellar dust. In the following
analysis we consider the 60\,$\mu$m emission as originating
from orbiting circumstellar dust, but note that far-infrared imaging is necessary
to concretely identify the origin of this star's far-infrared excess emisson.

Dust temperature and fractional infrared luminosity estimates for the
material orbiting HD\,121191 are performed in a similar manner to 
HD\,131488 (Section \ref{sec131488irex}). The blackbodies used
in fitting the excess emission seen toward HD\,121191 have
temperatures of 450 and 95\,K (Figure \ref{fig121191sed})
and result in a total fractional infrared luminosity
of L$_{\rm IR}$/L$_{\rm star}$$\approx$4.9$\times$10$^{-3}$.
Decomposing this into the hot and cold dust components it
is found that the hot dust contributes $\approx$2.3$\times$10$^{-3}$
to the fractional infrared luminosity while the cold component
contributes $\approx$2.6$\times$10$^{-3}$. From the spectral
energy distribution fit T$_{\rm star}$$=$7700\,K and the derived
L$_{\rm star}$$=$7.9\,L$_{\odot}$ we determine that blackbody
dust grains having temperatures of 450 and 95\,K would have
orbital semi-major axes of $\approx$1.3 and $\approx$28\,AU. Similar to the case
of HD\,131488, and given the question of the origin of the far-infrared
excess emission, it is possible that these values will change
when complete far-infrared measurements become available.

\section{Discussion}
\label{secdisc}

Given the young ages of HD\,131488 and HD\,121191, 
it is prudent to wonder whether their orbiting material 
could be remnant protoplanetary disk material left over from the
star formation event. If this were the case, and given the location of the
inner planetary system dust (Sections \ref{sec131488irex} and \ref{sec121191irex}), 
we would expect to see some gaseous material infalling onto the star that would 
manifest itself as emission lines in an optical spectrum. We see no evidence in either our
low- or high-resolution spectra for
infalling gaseous material (Figure \ref{fig131488dbs}).
When considered with the results of \citet{zuckerman95a}, \citet{carpenter06},
and \citet{currie09} regarding the more
rapid evolution of disks around early-type stars, the lack
of signatures of gas accreting onto either star
leads us to conclude that their orbiting material is
second generation debris that is being released by the collisions of mature
rocky objects.

Modeling the observed excess infrared emission shows that 
both HD\,131488 and HD\,121191 host dusty debris
in regions of their planetary system with temperature-equivalent separations
in our Solar system comparable to the orbit of Mercury
and close to that of Saturn. For the cool dust in each system we estimate
that the maximum fractional infrared luminosity that could be produced
by steady-state collisions of planetesimals
\citep[f$_{\rm max}$ in Equation 18 from][]{wyatt08} is $\sim$0.1\%. Since
the cool dust fractional infrared luminosity
(L$_{\rm IR}$/L$_{\rm bol}$; hereafter $\tau$) 
for both HD\,131488 and HD\,121191 is not significantly
greater than the f$_{\rm max}$ value, this
suggests that the cool dust orbiting both stars is readily explained as
originating in steady-state collisions among rocky objects
in a Kuiper-belt analog \citep{wyatt08}.

HD\,131488 and HD\,121191 are the first early-type stars identified with hot debris
disks having large fractional infrared luminosities and substantial outer debris belts.
To date, other early type stars, when hosting copious amounts of inner planetary system
dust, showed no signs of dust production in their outer planetary systems
(see Table \ref{tabamix} and references therein). Thus,
HD\,131488 and HD\,121191 join a growing class of objects previously only populated by
Sun-like stars $-$ namely V488\,Per \citep{zuckerman12} and HD\,166191
\citep{schneider13} $-$
to host both substantial inner and outer planetary system debris belts. It is worth noting
that a strong solid-state emission feature detected toward HD\,166191 almost
certainly originates from silicate species \citep{schneider13},
indicating that the unusual feature potentially seen in the mid-infrared spectra of
HD\,131488 and HD\,121191 is not a common feature of stars hosting both inner
and outer planetary system dust.
V488\,Per, despite being the dustiest main sequence star known to date
and host to hot and cold dust, has no mid-infrared spectrum that we know of.

The hot dust components seen for each star host $\tau$ values that are the largest
yet seen for A-type stars. Table \ref{tabamix} compiles all main sequence 
stars of spectral type earlier than F3 which host confirmed inner planetary system material with
$\tau$$\gtrsim$10$^{-3}$. It is clear that HD\,131488 and HD\,121191 dominate other
stars of spectral type A. Also notable is a dramatic drop in the most extreme $\tau$
values observed when crossing from spectral types F to A. Of special interest, though,
is the tight clustering of the ages of these sources. Even when one considers the range of
plausible ages for these systems $-$ which is on the order of 10-20\,Myr for all Table \ref{tabamix}
stars except HD\,145263 (whose age is likely between 3-10\,Myr) $-$ there are no systems
with ages above 20\,Myr. This result is despite numerous targeted surveys and blind 
cross-correlation
searches that included early-type stars with ages greater than 20\,Myr (e.g., \citealt{rieke05},
\citealt{moor06}, \citealt{su06}, \citealt{rhee07a}, \citealt{currie08a}, \citealt{balog09}, \citealt{melis09phd}, \citealt{morales09}, \citealt{zuckerman11}, \citealt{chen12}).
This age
clustering suggests that early-type stars experience large dust-producing events at an earlier
time than do Sun-like stars which have been shown to host extremely dusty systems
between ages of 30-100\,Myr \citep{melis10}. If these dust-producing events
at early-type stars 
are associated with terrestrial planet building as was suggested for the Sun-like stars
in \citet{melis10}, then the indication is that early-type stars complete the terrestrial planet
building process much more rapidly than do stars like the Sun.

\begin{figure}
 \centering
 \includegraphics[width=130mm]{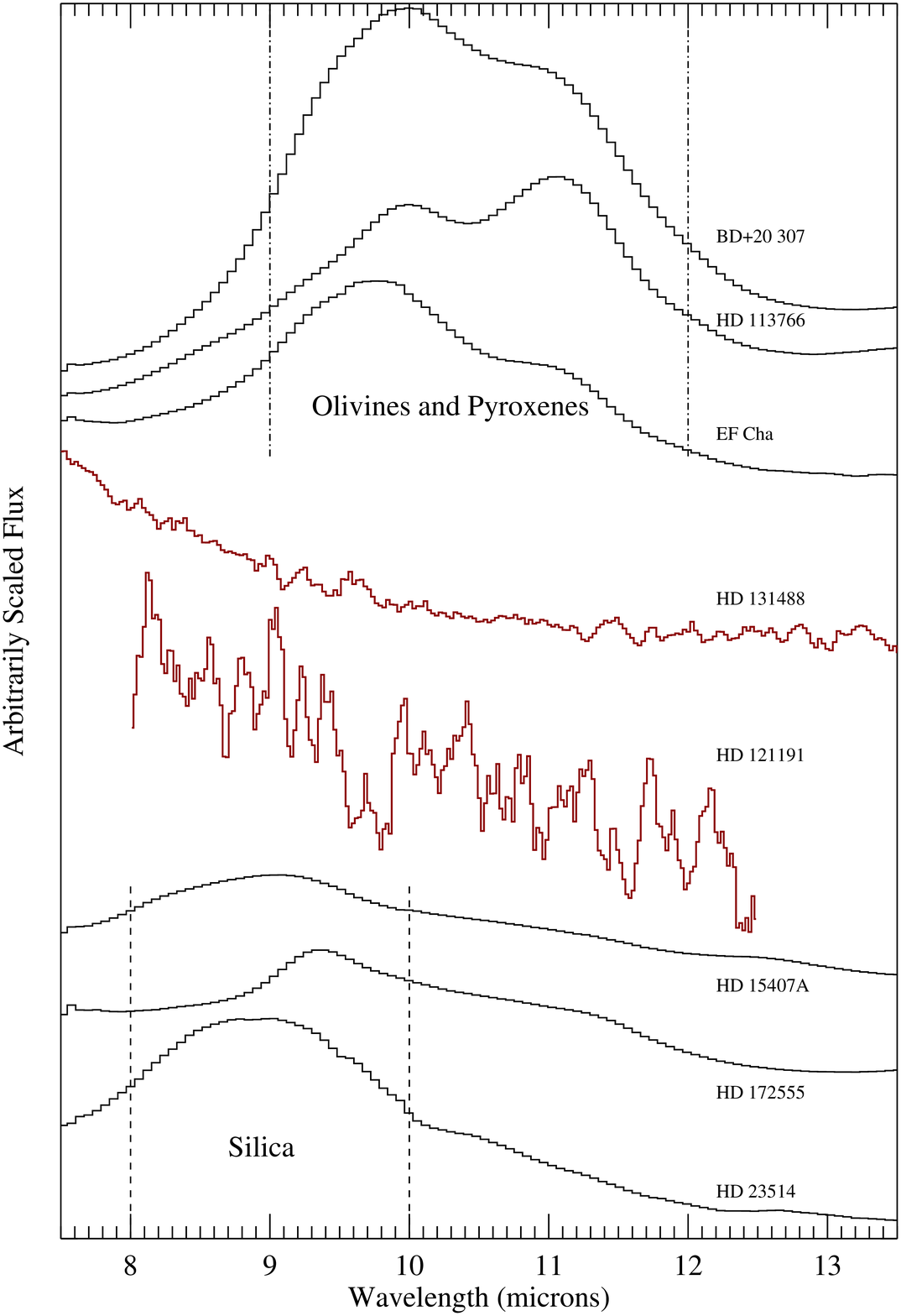}
 \caption{\label{fig131488speccomp} Comparison of HD\,131488's and HD\,121191's 
                T-ReCS spectra
                with silicate emission feature exemplars observed with IRS
                (note that the IRS spectra have been heavily smoothed to highlight their major
                features. 
                BD+20~307: \citealt{weinberger11}; HD\,113766: \citealt{chen06}, \citealt{lisse08};
                EF\,Cha: \citealt{currie11}; HD\,15407A: \citealt{melis10}, \citealt{fujiwara12};
                HD\,172555: \citealt{chen06}, \citealt{lisse09};
                HD\,23514: \citealt{rhee08}, \citealt{melis10}). 
                The signal-to-noise ratio is low
                for HD\,121191 (Table \ref{tabirspec}), 
                but the overall slope similar to that of HD\,131488 is apparent.
                An emission feature 
                that peaks at wavelengths shortward of 8\,$\mu$m in the T-ReCS
                mid-infrared spectra cannot be explained with
                typically seen minerals (see Section \ref{secdisc}).}
\end{figure}

The strange profile detected in the T-ReCS N-band
spectra adds an extra peculiarity to the cases of HD\,131488 and HD\,121191. The SpeX data
and the T-ReCS N-band photometric measurement for HD\,131488
require that the red wavelengths
of the T-ReCS N-band spectrum be roughly consistent with the dashed model sum
line plotted in Figure \ref{fig131488sed}. 
When the red part of the T-ReCS N-band spectrum is anchored as such, then 
the rising slope at blue
wavelengths of the T-ReCS spectrum must turn over to meet the red end of the
SpeX spectrum as it rises off of the stellar photosphere. These features
suggest an emission ``bump'' in the mid-infrared that is likely due to a broad solid-state
resonance in emission. The T-ReCS spectrum for HD\,121191, although of low
signal-to-noise ratio, presents a slope like that seen in
HD\,131488 (Figure \ref{fig131488speccomp}) and hence suggests a similar feature
in its mid-infrared spectrum.
Comparing to other well-studied sources (Figure \ref{fig131488speccomp})
suggests that the unusual solid state emission feature cannot be due to more commonly
observed minerals like olivine, pyroxene, or silica 
\citep[e.g.,][]{chen06,lisse08,lisse09,sargent09b}. 
Unfortunately, the limited data set in hand prevents conclusive analysis. We note
that carbonaceous materials, like calcite (CaCO$_3$), are able to reproduce
reasonably well an emission feature with peak near 7\,$\mu$m and the red shoulder
of the T-ReCS N-band spectra. But, these materials
often have other transitions that should be seen in our T-ReCS or SpeX spectra 
(e.g., the 11.5\,$\mu$m calcite band).
High quality mid-infrared spectroscopy
should be able to settle the existence and identity of any spectral features.
Until the hot dust mineralogy is settled, it cannot be concluded with
confidence that either of HD\,131488 or HD\,121191 have inner planetary system dust
generated by collisions between rocky bodies. For example,
one might consider a scenario where the mid-infrared continuum and emission feature is
generated by dust and water from a swarm of out-gassing cometary bodies
(C.\ Lisse 2012, private communication $-$ evaporating cometary bodies have 
been suggested as an explanation of optical and ultraviolet
absorption line variability in some dusty stars; see e.g., \citealt{ferlet87}; \citealt{montgomery12}).

\section{Conclusions}

We have identified two early-type main sequence stars that are orbited by copious 
amounts of hot dust at semi-major axes comparable to that of the Sun's
terrestrial planets.
In addition to hot inner planetary system material, both HD\,131488 and HD\,121191 are 
host to cooler dust in their outer planetary system and strange
solid-state emission features that make ambiguous the origin of their inner planetary
system dust.

Synthesis of those early-type stars (spectral types earlier than F3) with hot inner planetary
system dust and fractional infrared luminosities $\gtrsim$10$^{-3}$ suggests that these systems
are typically only found when their host stars have ages between 5-20\,Myr. If these
systems are linked to the final giant impact-type phase of terrestrial planet-building
as suggested for Solar-mass stars \citep{melis10}, then early-type stars would appear to
finish this stage of planet formation much sooner than do Sun-like stars (for which exceptionally
dusty systems have been observed in the age range of 30-100\,Myr; \citealt{melis10}).

\acknowledgments

We thank Alan Tokunaga for granting us IRTF/SpeX engineering time and
John Rayner for observing HD\,131488 for us. We thank Andrew Boden for
observing HD\,131488 with Keck~II/NIRC2-AO for us.
We are grateful to the Director of the Gemini South telescope for granting us
observing time.
We thank Carey Lisse for useful discussion. We thank the referee for a prompt review.
Based on observations obtained at the Gemini Observatory, which is operated by the
Association of Universities for Research in Astronomy, Inc., under a cooperative agreement
with the NSF on behalf of the Gemini partnership: the National Science Foundation (United
States), the Science and Technology Facilities Council (United Kingdom), the
National Research Council (Canada), CONICYT (Chile), the Australian Research Council
(Australia), MinistŽrio da Cincia e Tecnologia (Brazil) 
and Ministerio de Ciencia, Tecnolog'a e Innovaci—n Productiva  (Argentina).
Some of the data presented herein were obtained at the W.M. Keck Observatory, 
which is operated as a scientific partnership among the California Institute of 
Technology, the University of California and the National Aeronautics and Space 
Administration. The Observatory was made possible by the generous financial 
support of the W.M. Keck Foundation.
This publication makes use of data products from the Two Micron All Sky
Survey, which is a joint project of the University of Massachusetts and the
Infrared Processing and Analysis Center/California Institute of Technology,
funded by the National Aeronautics and Space Administration and the National
Science Foundation. This research made use of the SIMBAD and VizieR
databases.
C.M. acknowledges support from a LLNL Minigrant to UCLA. This
work was supported in part by the National Science Foundation under
award No.\ AST-1003318. 
This research was supported in part by NASA grants to
UCLA and University of Georgia.



{\it Facilities:} \facility{IRAS ()}, \facility{AKARI ()}, \facility{Gemini:South (T-ReCS)}, \facility{Keck:I (HIRES)}, \facility{Siding Spring Observatory 2.3-m (Echelle, Double Beam Spectrograph, WiFeS)}, \facility{Keck:II (NIRC2)}


\clearpage





\clearpage

\clearpage

\clearpage

\begin{deluxetable}{lcccccc}
\rotate
\tabletypesize{\normalsize}
\tablecolumns{7} 
\tablewidth{0pt}
\tablecaption{Mid-infrared Imaging Observations Summary \label{tabirim}}
\tablehead{
 \colhead{Star} &
 \colhead{UT Date} & 
 \colhead{Gemini Program} &
 \colhead{Instrument} & 
 \colhead{Filter} & 
 \colhead{Filter} & 
 \colhead{Photometric} \\ 
 \colhead{} &
 \colhead{} & 
 \colhead{} &
 \colhead{} & 
 \colhead{Name} & 
 \colhead{Wavelength ($\mu$m)} & 
 \colhead{Standard}
}
\startdata
HD 131488 & 16 Jul 2007 & GS-2007A-Q-43 & T-ReCS & N-broad & 10.36$\pm$5.3  & HD 130157  \\ 
HD 131488 & 06 Jun 2009 & GS-2009A-C-1 & T-ReCS &  N-broad & 10.36$\pm$5.3 & $-$\tablenotemark{a}  \\
HD 131488 & 06 Jun 2009 & GS-2009A-C-1 & T-ReCS &  Qa & 18.30$\pm$1.5  & $-$\tablenotemark{a}  \\
HD 121191 & 12-14 Aug 2010 & GS-2010B-DD-1 & T-ReCS &  N-broad & 10.36$\pm$5.3 & HD 119193\tablenotemark{b}  \\ 
\enddata
\tablenotetext{a}{Conditions were not photometric.}
\tablenotetext{b}{Photometry was
performed with spectroscopy acquisition images obtained 14 Aug 2010.}
\end{deluxetable}

\clearpage

\begin{deluxetable}{lccccc}
\tabletypesize{\small}
\tablecolumns{6}
\tablewidth{0pt}
\tablecaption{Broadband Photometry \label{tabphot}}
\tablehead{
 \colhead{Filter Name} &
 \colhead{$\lambda$} &
 \multicolumn{2}{c}{HD\,131488} &
 \multicolumn{2}{c}{HD\,121191} \\
 \colhead{} &
 \colhead{($\mu$m)} &
 \colhead{magnitude} &
 \colhead{Flux (mJy)} &
 \colhead{magnitude} &
 \colhead{Flux (mJy)}
}
\startdata
$B$ & 0.44 & 8.09$\pm$0.02 & 2323$\pm$43 & 8.40$\pm$0.02 & 1746$\pm$32 \\
$V$ & 0.55 & 8.00$\pm$0.01 & 2271$\pm$21 & 8.16$\pm$0.01 & 1960$\pm$18 \\
$J$ & 1.25 & 7.85$\pm$0.03 & 1180$\pm$33 & 7.73$\pm$0.02 & 1319$\pm$24 \\
$H$ & 1.65 & 7.85$\pm$0.05 & 761$\pm$35 & 7.68$\pm$0.02 & 890$\pm$16 \\
$K_{\rm s}$ & 2.20 & 7.80$\pm$0.03 & 506$\pm$14 & 7.70$\pm$0.02 & 555$\pm$10 \\
$W1$ & 3.35 & 7.72$\pm$0.02 & 250$\pm$5 & 7.59$\pm$0.03 & 282$\pm$8 \\
$W2$ & 4.60 & 7.56$\pm$0.02 & 161$\pm$3 & 7.54$\pm$0.02 & 164$\pm$3 \\
$AKARI$ IRC & 9 & $-$ & 164$\pm$7 & $-$ & 134$\pm$14 \\
T-ReCS N & 10.6 & $-$ & 121$\pm$12 & $-$ & 117$\pm$12 \\
$W3$ & 11.56 & 6.08$\pm$0.01 & 107$\pm$1 & 6.15$\pm$0.02 & 101$\pm$2 \\
$IRAS$ & 12 & $-$ & 169$\pm$30 & $-$ & 191$\pm$31 \\
$AKARI$ IRC & 18 & $-$ & $-$ & $-$ & 135$\pm$3 \\
$W4$ & 22.09 & 4.31$\pm$0.02 & 156$\pm$3 & 4.14$\pm$0.02 & 183$\pm$3 \\
$IRAS$ & 25 & $-$ & 151$\pm$33 & $-$ & 221$\pm$30 \\
$IRAS$ & 60 & $-$ & 348$\pm$63 & $-$ & 476$\pm$125 \\
\enddata
\tablecomments{$BV$ data are from Tycho-2 \citep{hog00} and are converted to the Johnson
system following \citet{bessell00}. $JHK_{\rm s}$ data are from 2MASS \citep{cutri03}. $W1W2W3W4$
data are from $WISE$ \citep{cutri12} and are not color corrected. $AKARI$ data are from the IRC all-sky catalog \citep{ishihara10}; HD\,131488 does not have a reported 18\,$\mu$m flux density in that catalog. $IRAS$ data are from the Faint Source Catalog \citep{moshir92} for
HD\,131488 and the Faint Source Reject Catalog for HD\,121191. Neither source was detected
by $IRAS$ at 100\,$\mu$m and the upper limits are not reported here as they are not restrictive (see Figure \ref{fig121191sed}).}
\end{deluxetable}

\clearpage

\begin{deluxetable}{lccccccc}
\rotate
\tabletypesize{\normalsize}
\tablecolumns{8} 
\tablewidth{0pt}
\tablecaption{Optical Spectroscopic Observations Summary \label{taboptspec}}
\tablehead{
 \colhead{Star} &
 \colhead{UT Date} & 
 \colhead{Instrument} & 
 \colhead{Setup} & 
 \colhead{Coverage (\AA )} & 
 \colhead{Resolution\tablenotemark{a}} & 
 \colhead{S/N} & 
 \colhead{$\lambda$ of S/N\tablenotemark{b} (\AA ) }
}
\startdata
HD 131488 & 14 Feb 2008 & HIRES & UV Collimator & 3000-5900 \AA\ & 40,000        & 100 & 4600 \AA\ \\
HD 131488 & 15 Feb 2008 & SSO-E & $-$                      & 3910-6730 & 25,000 & 50 & 6100 \\
HD 131488 & 26 Feb 2008 & HIRES & Red Collimator & 4600-9200 \AA\ & 40,000 & 100 & 6350 \AA\ \\
HD 131488 & 14 Jun 2008 & SSO-E & $-$                      & 3910-6730 & 25,000 & 70 & 6100 \\
HD 131488 & 17 Jun 2008 & DBS  & 300B, 3$^{\prime\prime}$ slit & 3300-6200 & 6.8 \AA\ & 100 & 4600 \\ 
HD 131488 & 13 Jul 2008 & SSO-E & $-$                        & 3910-6730 & 25,000 & 80 & 6100  \\
HD 131488 & 14 Jul 2008 & DBS  & 600B, 3$^{\prime\prime}$ slit & 3800-5500 & 3.3 \AA\ & 100 & 4600 \\
HD 131488 & 15 Jul 2008 & DBS  & 600R, 3$^{\prime\prime}$ slit & 5500-7300 & 3.3 \AA\ & 100 & 6700 \\
HD 121191 & 06 Jun 2010 & WiFeS & B$_{3000}$ & 3400-5900 &  3,000 & 100 & 4600 \\
HD 121191 & 24 Jun 2010 & SSO-E & $-$                        & 4130-6930 & 25,000 & 70 & 6100 \\
HD 121191 & 20 Jul 2010 & SSO-E & $-$                          & 3910-6720 & 25,000 & 100 & 6100 \\
\enddata
\tablenotetext{a}{Resolution measurements are from the FWHM of single arclines in our comparison spectra.}
\tablenotetext{b}{Wavelength where S/N measurement was made in the spectrum.}
\end{deluxetable}

\clearpage

\begin{deluxetable}{lcccccccc}
\rotate
\tabletypesize{\normalsize}
\tablecolumns{10} 
\tablewidth{0pt}
\tablecaption{Infrared Spectroscopic Observations Summary \label{tabirspec}}
\tablehead{
 \colhead{Star} &
 \colhead{UT Date} & 
 \colhead{Instrument} & 
 \colhead{Setup} & 
 \colhead{Coverage} & 
 \colhead{Resolution} & 
 \colhead{Telluric} &
 \colhead{S/N} & 
 \colhead{$\lambda$ of S/N\tablenotemark{a}} \\
 \colhead{} &
 \colhead{} & 
 \colhead{} & 
 \colhead{} & 
 \colhead{($\mu$m)} & 
 \colhead{} & 
 \colhead{Standard} &
 \colhead{} & 
 \colhead{($\mu$m)}
}
\startdata
HD 131488 & 27 Apr 2009 & T-ReCS & 0.7$''$ slit & 7.5-13.5  & 100 & HD 130157 & 20 & 10.5 \\
HD 131488 & 02 Jul 2009 & SpeX       & 0.5$''$ slit & 2.2-5.0 & 1500 & HD 130163 & 20 & 4  \\
HD 121191 & 14 Aug 2010 & T-ReCS & 0.26$''$ slit & 7.5-13.5 & 300 & HD 119193 & 7  & 10.5 \\
\enddata
\tablecomments{HD\,131488 Gemini data are from
GS-2009A-DD-5 while HD\,121191 data are from GS-2010B-DD-1.}
\tablenotetext{a}{Wavelength where signal-to-noise ratio measurement was made in the spectrum.}
\end{deluxetable}

\clearpage

\begin{deluxetable}{lcc}
\tabletypesize{\small}
\tablecolumns{3}
\tablewidth{0pt}
\tablecaption{Stellar Parameters\label{tabpars}}
\tablehead{
 \colhead{Parameter} &
 \colhead{HD\,131488} &
 \colhead{HD\,121191}
}
\startdata
R.A.\ (J2000) & 14 55 08.03 & 13 55 18.89 \\
Decl.\ (J2000) & $-$41 07 13.3 & $-$54 31 42.7 \\
V$_{\rm mag}$ & 8.00 & 8.16 \\
B$-$V & 0.09 & 0.24 \\
Sp.\ Type & A1~V & A5~IV/V  \\
T$_{\rm star}$ (K) & 8800 & 7700 \\
R$_{\rm star}$ (R$_{\odot}$) & 1.7 & 1.6 \\
L$_{\rm star}$ (L$_{\odot}$) & 15.5 & 7.9 \\
pmRA (mas yr$^{-1}$)& $-$19.0$\pm$1.3 & $-$26.7$\pm$1.4 \\
pmDE (mas yr$^{-1}$) & $-$21.8$\pm$1.2 & $-$18.0$\pm$1.3 \\
RV (km s$^{-1}$) & +7$\pm$2 & +12$\pm$3 \\
Dist.\ (pc) & $\approx$150 & $\approx$130 \\
UVW (km s$^{-1}$) & $-$4, $-$21, $-$5 & $-$6, $-$23, $-$6 \\
Assoc.\ & UCL & LCC\tablenotemark{c} \\
\enddata
\tablecomments{\footnotesize{R.A.\ and Decl.\ are from 2MASS \citep{cutri03}. 
V$_{\rm mag}$ and B$-$V are from Tycho-2 \citep{hog00}; see note to Table \ref{tabphot}. 
Spectral types are from the SIMBAD database.
T$_{\rm star}$ and R$_{\rm star}$ are extracted from atmospheric model fits to broadband
photomteric measurements. L$_{\rm star}$ is then calculated using L$=$4$\pi$R$^{2}$$\sigma$$_{\rm SB}$T$^{4}$.
pmRA and pmDE are from Tycho-2. 
Radial velocities are measured from our high- and medium-resolution echelle spectra.
See Section \ref{secres} for how distances are estimated.
UVW for the UCL and LCC subgroups are roughly
$-$7\,km\,s$^{-1}$, $-$20\,km\,s$^{-1}$, and $-$6\,km\,s$^{-1}$ \citep{sartori03,chen11}.} Uncertainties for UVW in the table are typically 2-3 km s$^{-1}$.}
\tablenotetext{c}{HD\,121191 lies between the UCL and LCC associations in the plane of the sky and could potentially belong to either group.}
\end{deluxetable}

\clearpage

\begin{deluxetable}{lcrcccccc}
\rotate
\tabletypesize{\footnotesize}
\tablecolumns{9}
\tablewidth{0pt}
\tablecaption{Confirmed Main Sequence A- and Early F-type Stars with Hot Dust Luminosities $\gtrsim$10$^{-3}$\,L$_{\rm bol}$ \label{tabamix}}
\tablehead{ 
  \colhead{Star} &
  \colhead{Spectral} &
  \colhead{Age} &
  \colhead{T$_{dust}$} &
  \colhead{$\tau$$_{\rm hot}$\tablenotemark{a}} &
  \colhead{R$_{\rm dust}$} &
  \colhead{Cool} &
  \colhead{Solid-State} &
  \colhead{References} \\
  \colhead{} &
  \colhead{Type\tablenotemark{b}} &
  \colhead{(Myr)} &
  \colhead{(K)} &
  \colhead{($\times$10$^{-4}$)} &
  \colhead{(AU)} &
  \colhead{Dust?} &
  \colhead{Features?\tablenotemark{c}} &
  \colhead{}
}
\startdata
 HD 131488 & A2 & 10 & 750,100 & 40\tablenotemark{d} & 0.6 & Y & Y & 1 \\
 HD 172555 & A7 & 12 & 320 & 8.1 & 2.6 & N & Y & 2,3,4 \\
 HD 121191 & A8 & 10 & 450,95 & 23\tablenotemark{e} & 1.2 & Y? & Y & 1 \\
 EF\,Chamaeleontis & A9 & 10 & 240 & 10 & 4.9 & N & Y & 3,5 \\
 HD 113766 & F0 & 10 & 350 & 150 & 1.9 & N & Y & 2,6,7 \\
 HD 145263 & F2 & 5 & 200 & $\sim$100 & 5.2 & N & Y & 8,9 \\
\tableline
\enddata
\tablerefs{(1) this work, (2) \citet{chen06}, (3) \citet{rhee07b}, (4) \citet{lisse09},
(5) \citet{currie11}, (6) \citet{lisse08}, (7) \citet{olofsson12}, 
(8) \citet{honda04}, (9) \citet{chen11}. }
\tablenotetext{a}{$\tau$$=$L$_{\rm IR}$/L$_{\rm bol}$, the fractional infrared luminosity of the dust.}
\tablenotetext{b}{Spectral types are estimated from temperatures obtained when fitting atmospheric models to the stellar spectral energy distribution.}
\tablenotetext{c}{Y indicates that solid-state emission features have been detected in mid-infrared spectra of the source.}
\tablenotetext{d}{$\tau$$_{\rm total}$ = 60 $\times$ 10$^{-4}$, see Section \ref{sec131488irex}.}
\tablenotetext{e}{$\tau$$_{\rm total}$ = 49 $\times$ 10$^{-4}$, see Section \ref{sec121191irex}.}
\end{deluxetable}






\end{document}